\documentclass[prb,twocolumn,english]{revtex4}
\usepackage[pdftex]{graphicx}
\usepackage{bm}
\usepackage{epstopdf}
\usepackage{amsmath}

\begin{document}

\title{Hydrodynamic Collective Effects of Active Protein Machines in Solution and Lipid Bilayers}
\author{Alexander S. Mikhailov}
\affiliation{Abteilung Physikalische Chemie, Fritz-Haber-Institut der Max-Planck-Gesellschaft, Faradayweg 4-6, 14195 Berlin, Germany}
\author{Raymond Kapral}
\affiliation{Chemical Physics Theory Group, Department of Chemistry, University of Toronto, Toronto, Ontario M5S 3H6, Canada}

\begin{abstract}
The cytoplasm and biomembranes in biological cells contain large numbers of proteins that cyclically change their shapes. They are molecular machines that can function as molecular motors or carry out many other tasks in the cell. We analyze the effects that hydrodynamic flows induced by active proteins have on other passive molecules in solution or membranes. We show that the diffusion constants of passive particles are enhanced substantially. Furthermore, when gradients of active proteins are present, a chemotaxis-like drift of passive particles takes place. In lipid bilayers, the effects are strongly nonlocal, so that active inclusions in the membrane contribute to diffusion enhancement and the drift. The results indicate that the transport properties of passive particles in systems containing active proteins machines operating under nonequilibrium conditions differ from their counterparts in systems at thermal equilibrium.
\end{abstract}

\maketitle

\section{Introduction}
Protein machines play a fundamental role in biological cells.~\cite{alberts-cell,jones-book} Operating as motors, they are responsible for intracellular transport and force generation. As manipulators, they perform various operations involving other biomolecules, including RNA and DNA. As pumps, they transfer ions across biomembranes. A common feature of protein machines is that they undergo cyclic conformational changes that are induced by ligand binding and product release. Thus, protein machines are enzymes where substrate binding, catalytic conversion to products and product release are accompanied by internal mechanochemical motions. Conformational changes within turnover cycles are also characteristic of many other enzymes, which need not function as molecular machines. The results we present in this paper are also applicable to such enzymes.

When a macromolecule cyclically changes its shape, it induces hydrodynamic flows in the surrounding fluid or biomembrane in which it resides. Such pulsating flows can act on other passive particles in solution or lipid bilayers. The aim of the present study is to analyze the collective hydrodynamic effects that active macromolecules have on passive particles in the medium. We shall show that these effects lead to substantial modifications of the diffusion constants of passive particles. Furthermore, directed drift of passive particles can be induced when there are spatial gradients of active macromolecules, a phenomenon that is reminiscent of chemotaxis.

The investigation of hydrodynamic effects in active fluids is an important field of current research.~\cite{lauga:09,goldstein:07} While the hydrodynamics of bacterial motion has been studied often, the active elements may be of inorganic origin and operate through various flow-generation mechanisms.~\cite{wangbook:13,SenRev:13,kapral:13,Seifert_12,ozin:10,golestanian:07_2,mikhailov-cells,julicher:97,aranson:06} There has been a considerable amount of work on  swimmers that can propel themselves by cyclically changing their shapes.~\cite{purcell:77} Interactions between such swimmers and their collective flows have been analyzed.~\cite{saintillan:12,poon:14,stark:14,sriram-review:13,bocquet:12,speck:13,golestanian:14} Also, interactions between active hydrodynamic dipoles have been investigated theoretically~\cite{yeomans:08,lauga:08} and experimentally.~\cite{shani:14,goldstein:11}

Hydrodynamic effects on individual protein machines have been studied; for example, investigations of simple models of such machines have shown how they propel themselves and behave under a load~\cite{sakaue:10}; also, the effects of hydrodynamic interactions on the internal dynamics have been analyzed~\cite{CTMK08}. Active protein inclusions in lipid bilayers can act as hydrodynamic dipoles~\cite{huang:12} and, under certain conditions, such inclusions can behave as active membrane swimmers~\cite{huang:12b}.

The focus of the present study differs in several respects from the work recounted above. We are not be interested in the effects of hydrodynamics on the operation of a single machine, nor do we consider interactions between protein machines and the collective flows exhibited by such hydrodynamically-coupled active molecules. Instead, we concentrate on the effects that protein machines can have on passive particles in the medium. While some proteins can indeed behave as motors, we only require that such machines act as cyclic hydrodynamic dipoles. Consequently, our analysis concerns the statistical effects that populations of incoherently oscillating dipoles can have on passive particles in the system.

In bulk solution, the laws of three-dimensional (3D) hydrodynamics need to be applied; however, as already pointed out by Saffmann and Delbr\"uck~\cite{Saffman_Delbruck_75}, biological membranes should behave as two-dimensional (2D) fluids when lipid flows in a membrane that occur on scales shorter than a micrometer are considered.~\cite{Diamant_09,Saffman_Delbruck_75,DiLeonardo:08} Recently, 2D lipid flows were directly observed in mesoscopic simulations of lipid bilayers~\cite{huang:13}; additionally, 2D diffusion in biomembranes was experimentally investigated~\cite{cheung:96}. It is well known that 2D hydrodynamics is characterized by the presence of ultra-long-ranged logarithmic interactions that make it qualitatively different from the 3D case.~\cite{DiLeonardo:08} We study the hydrodynamic effects of active machines in both 3D and 2D systems.

The outline of the paper is as follows. First, we present a simple model of a protein machine as a dimer undergoing random cyclic changes. This model is only used as an illustration and our general results do not rely on this specific structure of the model. We then show that a cyclically fluctuating hydrodynamic dipole will induce diffusive motion and directed drift of a passive particle located at some distance from it. When a population of cyclic hydrodynamic dipoles is randomly distributed in medium, they will enhance the diffusion of all passive particles in the medium. Moreover, if such dipoles are non-uniformly distributed and concentration gradients in these species are present, directed flows of all passive particles will be induced. Numerical estimates of the magnitudes of the effects are given, and a discussion of the results is presented.

\section{Protein machines as force dipoles}

Molecular machines are biomolecules, most often proteins, that undergo structural changes in shape during their operation cycles. These shape changes, induced by ligand binding and product release, take place under nonequilibrium conditions; therefore, they differ from thermally-induced shape fluctuations for which microscopic reversibility holds and the fluctuation-dissipation theorem applies. These molecular machines operate in a viscous environment and their dynamics takes place under low Reynolds number conditions so that inertia does not play a significant role. As a result, if a force is applied to a particle in the fluid, the same force acts on the fluid. More specifically, such protein machines act as stochastic oscillating force dipoles that can influence the motions of other particles in the system.

In order to illustrate how oscillating force dipoles arise in protein machines, consider a coarse-grain model of a protein comprising two domains, represented by beads, and connected by an elastic spring. If the spring contracts or stretches, two forces – equal in magnitude and opposite in the direction – will act on the beads, and also on the surrounding fluid, creating a force dipole of strength $m=Fl$ , where $F$  is the magnitude of the force and $l$  is the distance between the beads.   As a consequence of ligand binding and product release the characteristics of the spring force will change.

A simple model for such ligand-induced force changes is as follows. The overdamped motion of the beads, labeled by 1 and 2, is given by $\dot{x}_1=\mu k (x_2-x_1-l(s))$, where $\mu$  is the mobility of the beads, $k$  is the stiffness constant of the elastic spring, and $l(s)$  is the natural spring length which depends on the variable $s=\{0,1\}$  that specifies the binding state of the model machine:  $l(s=0)=a$ and $l(s=1)=b$ . If the binding state of the machine is fixed, it reaches a steady configuration with the spring length $a$ or $b$ and remains indefinitely in that state. Suppose however that the internal state can change as a result of reactions, so that stochastic transitions from $s = 0$ to $s = 1$ and back take place. Generally, such transitions will depend on the length of the spring: close to the state with $l=a$ the transition from $s = 0$ to $s = 1$ becomes possible, whereas near the state with $l=b$  the reverse transition can occur. Under these conditions, the machine will go through cycles of expansion and contraction, similar to the cyclic conformational changes in a real protein machine. It is an oscillating force dipole which can act on the surrounding viscous fluid to create hydrodynamic flows that can induce motions in other passive particles

\section{Hydrodynamic effects}
When a force ${\bf F}$ is applied to the fluid at a point ${\bf r}$, it produces a fluid flow field at ${\bf R}$ that advects a particle at this location with velocity,
\begin{equation}\label{eq:oseen-ap}
\dot{R}_\alpha= G_{\alpha \beta}({\bf R}-{\bf r}) F_\beta({\bf r}),
\end{equation}
where $G_{\alpha \beta}$  is the mobility tensor which, for sufficiently large distances, can be evaluated in the Oseen approximation. (The Einstein summation convention will be used throughout this paper.) For an oscillating dimer of length $x$ with orientation given by the unit vector ${\bf e}$ and interaction force F, we have $\dot{R}_\alpha= [G_{\alpha \beta}({\bf R}-{\bf r}-x {\bf e})- G_{\alpha \beta}({\bf R}-{\bf r})] e_\beta F$. If the dimer length $x$ is small compared to the distance $|{\bf R}-{\bf r}'|$ , we can write
\begin{equation}\label{eq:fd-ap}
\dot{R}_\alpha=\frac{\partial G_{\alpha \beta}}{\partial r_\gamma} e_\beta e_\gamma m,
\end{equation}
where $m(t)=x(t) F(t)$  is the force dipole.

Although we  derived eq.~(\ref{eq:fd-ap}) for a specific model, it is general. If an object immersed in the fluid changes its shape, it generates a hydrodynamic flow that, at large separations from this object, can be described as being produced by a force dipole. Below, we shall treat a protein machine as a stochastic force dipole m(t) with zero mean, $\langle m(t)\rangle_n=0$, and correlation function $S(t)=\langle m(t)m(0)\rangle_n$, where $\langle \cdots \rangle_n$ denotes an average over the stochastic fluctuations of the force dipole. 	According to eq.~(\ref{eq:fd-ap}), a force dipole will induce motion in a passive particle in the fluid.

Consider a collection of  such force dipoles in the system, located at positions $\{{\bf R}_i\}$ with orientations $\{{\bf e}_i\}$. If, within the time interval being considered, the displacements in the position ${\bf R}(t)$ of a passive particle are small, we can write $R_\alpha(t)=R_{0,\alpha}+\rho_\alpha(t)$  and retain only terms that are first order in the displacements ${\rho}(t)$ , so that
\begin{eqnarray}\label{eq:micro-rhodot}
\dot{\rho}_\alpha&=&\sum_{i}\Big[\frac{\partial G_{\alpha \beta}({\bf R}_{0}-{\bf R}_{i})}{\partial R_{i,\gamma}} +
\frac{\partial^2 G_{\alpha \beta}({\bf R}_{0}-{\bf R}_{i})}{\partial R_{i,\gamma} \partial R_{i,\delta}}  \rho_\delta \Big] \nonumber \\
&& \times e_{i,\beta} e_{i,\gamma} m_i(t).
\end{eqnarray}
It is convenient to write this equation in field-point notation,
\begin{eqnarray}\label{eq:micro-rhodot-field}
\dot{\rho}_\alpha&=&\sum_{i}\int d{\bf r}\;\Big[\frac{\partial G_{\alpha \beta}({\bf R}_0-{\bf r})}{\partial r_\gamma}  +
\frac{\partial^2 G_{\alpha \beta}({\bf R}_0-{\bf r})}{\partial r_\gamma \partial r_\delta}  \rho_\delta \Big] \nonumber \\
&&
\times \sum_{i}e_{i,\beta} e_{i,\gamma} m_i(t) \delta({\bf R}_{i}-{\bf r}),
\end{eqnarray}
where the last term represents the microscopic density of the product $e_{i,\beta} e_{i,\gamma} m_i(t) $ at ${\bf r}$. Note that the first and the second derivatives of the Green function correspond to dipole and quadrupole terms. This equation shows that the instantaneous position of a passive particle evolves with time according to a stochastic differential equation with multiplicative noise. We assume that the fluctuations of the different force dipoles are uncorrelated so that $\langle m_i(t)m_j(0)\rangle_n=S(t) \delta_{ij}$.

The diffusion constant $D$ and mean drift velocity $\overline{{\bf V}}$of the passive particle under the action of a field of fluctuating hydrodynamic dipoles can be determined from
\begin{equation}\label{eq:diff-drift}
D=\frac{1}{d} \int_0^\infty dt \; \langle \delta V_\alpha(t) \delta V_\alpha(0)\rangle, \quad \overline{V}_\alpha=\langle V_\alpha \rangle,
\end{equation}
where $d$ is the dimensionality of the system, $ \delta V_\alpha= V_\alpha -\overline{V}_\alpha$ and the angle bracket $\langle \cdots \rangle$ denotes an average over the stochastic fluctuations of the force dipoles, as well as their orientations and positions.

The expression in eq.~(\ref{eq:micro-rhodot-field}) for the velocity $V_\alpha=\dot{\rho}_\alpha$ of a particle may then be substituted into these equations, retaining only the leading terms, to obtain the diffusion coefficient and drift velocity. When computing the average values in eqs.~(\ref{eq:diff-drift}) we assume that the orientations of the force dipoles are not correlated with their positions so that $\langle  \sum_i  e_\beta e_{\beta'} e_\gamma e_{\gamma'} \delta({\bf R}_{i}-{\bf r})  \rangle =\Omega_{\beta \beta' \gamma \gamma'} c({\bf r})$, where $\Omega_{\beta \beta' \gamma \gamma'}=\langle e_\beta e_{\beta'} e_\gamma e_{\gamma'} \rangle$ and $c({\bf r})= \langle \sum_i \delta({\bf R}_{i}-{\bf r}) \rangle$ is the local concentration of force dipoles at point ${\bf r}$ in the fluid. We find
\begin{eqnarray}\label{eq:diff-gen}
&&D({\bf R}_0)=\frac{S }{d} \Omega_{\beta \beta' \gamma \gamma'}\\
&&\qquad \qquad \times \int d{\bf r}\; \frac{\partial G_{\alpha \beta}({\bf R}_0-{\bf r})}{\partial r_\gamma} \frac{\partial G_{\alpha \beta'}({\bf R}_0-{\bf r})}{\partial r_{\gamma'}} c({\bf r}),\nonumber
\end{eqnarray}
\begin{eqnarray}\label{eq:drift-gen}
&&\overline{V}_\alpha({\bf R}_0)= S \Omega_{\beta \beta' \gamma \gamma'} \\
&& \qquad \qquad \times \int d{\bf r}\; \frac{\partial^2 G_{\alpha \beta}({\bf R}_0-{\bf r})}{\partial r_\gamma \partial r_\delta} \frac{\partial G_{\delta \beta'}({\bf R}_0-{\bf r})}{\partial r_{\gamma'}} c({\bf r}).\nonumber
\end{eqnarray}
Here $S=\int_0^\infty dt \;S(t)$. We shall now analyse these equations separately for 3D and 2D systems.

\subsection{Three-dimensional systems}
For applications to protein machines in bulk solution, for example in the cytoplasm of biological cells, the 3D Green function in the Oseen approximation is $G_{\alpha \beta}({\bf r})=(8 \pi \eta r)^{-1}(\delta_{\alpha \beta}+ \hat{r}_\alpha \hat{r}_\beta)$  where $\eta$  is the fluid viscosity and $\hat{{\bf r}}$  is the unit vector specifying the direction of ${\bf r}$. Suppose that the machines are uniformly distributed in space and their average concentration is $c_0$. Then, eq.~(\ref{eq:diff-gen}) yields
\begin{equation}\label{eq:diff-3D}
D=\frac{S c_0}{d}\Omega_{\beta \beta' \gamma \gamma'} \int^{\prime} d{\bf r}\; \frac{\partial G_{\alpha \beta}({\bf r})}{\partial r_\gamma} \frac{\partial G_{\alpha \beta'}({\bf r})}{\partial r_{\gamma'}}.
\end{equation}
In writing this expression we have made the change of integration variable, ${\bf R}_0-{\bf r} \to {\bf r}$, retaining the same symbol to avoid proliferation of notation.
The prime on the integral indicates that the integral is cut off at short distances, since overlap between the passive particle and an active enzyme is restricted by volume exclusion. We denote this cut-off distance by $\ell_c$ and choose it to be larger than the mean diameter of the passive and active particles to account for the fact that the Oseen approximation breaks down at distances larger than the particle sizes. Introducing dimensionless coordinates ${\bf z}={\bf r}/\ell_c$, the diffusion coefficient can be estimated as
\begin{equation}\label{eq:diff-est}
D \approx \frac{S c_0}{\ell_c \eta^2} \zeta,
\end{equation}
where the positive factor $\zeta$ is given by
\begin{equation}\label{eq:zeta}
\zeta = \frac{\Omega_{\beta \beta' \gamma \gamma'}}{192 \pi^2} \int^{\prime} d{\bf z}\; \frac{\partial g_{\alpha \beta}({\bf z})}{\partial z_\gamma} \frac{\partial g_{\alpha \beta'}({\bf z})}{\partial z_{\gamma'}},
\end{equation}
and the dimensionless function $g_{\alpha \beta}({\bf z})=z^{-1}(\delta_{\alpha \beta} +\hat{z}_\alpha  \hat{z}_\beta)$.

Turning to eq.~(\ref{eq:drift-gen}), we notice that the drift velocity vanishes for a uniform distribution of active particles. Suppose instead that a constant concentration gradient in the direction $\hat{{\bf n}}$, ${\bm \nabla} c=\hat{{\bf n}}(\hat{{\bf n}}\cdot {\bm \nabla} c)= \hat{{\bf n}} (\nabla c)$, is present and  $c({\bf r})= c_0 +({\bf r}\cdot \hat{{\bf n}}) \nabla c$. Now, the integration in eq.~(\ref{eq:drift-gen}) yields
\begin{equation}
\overline{V}_\alpha=S\Omega_{\beta \beta' \gamma \gamma'} \int^{\prime} d{\bf r}\;  \frac{\partial^2 G_{\alpha \beta}({\bf r})}{\partial r_\gamma \partial r_\delta} \frac{\partial G_{\delta \beta'}({\bf r})}{\partial r_{\gamma'}}({\bf r}\cdot \hat{{\bf n}}) \nabla c.
\end{equation}
Again, introducing dimensionless coordinates ${\bf z}={\bf r}/\ell_c$, we find
\begin{equation}\label{eq:drift-est}
\overline{{\bf V}} \approx \frac{S}{\ell_c \eta^2} \xi {\bm \nabla} c,
\end{equation}
where the positive dimensionless factor $\xi$ is
\begin{equation}\label{eq:xi}
\xi = \frac{\Omega_{\beta \beta' \gamma \gamma'}}{64 \pi^2} \int^{\prime} d{\bf z}\; \hat{n}_\alpha\frac{\partial^2 g_{\alpha \beta}({\bf z})}{\partial z_\gamma \partial z_\delta} \frac{\partial g_{\delta \beta'}({\bf z})}{\partial z_{\gamma'}} ({\bf z}\cdot \hat{{\bf n}}).
\end{equation}

\subsection{Two-dimensional systems}
As noted earlier by Saffman and Delbr\"uck~\cite{Saffman_Delbruck_75}, biological membranes should behave as two-dimensional lipid fluids on sub-micrometer length scales. Therefore, the effects of an ensemble of active protein machines on the motion of a passive particle in a lipid bilayer is an example where a 2D description is appropriate. In the Oseen approximation, the 2D Green function of lipid bilayers is~\cite{Diamant_09}
\begin{equation}
G_{\alpha \beta}({\bf r})=(4 \pi \eta_m)^{-1}(-(1+\ln(\kappa r))\delta_{\alpha \beta}+ \hat{r}_\alpha \hat{r}_\beta)
\end{equation}
Here, $\eta_m$  is the two-dimensional viscosity of the lipid bilayer, which is related to its 3D viscosity $\eta_L$ by $\eta_m=\eta_L h$, where $h$ is the thickness of the bilayer. In contrast to the 3D case, hydrodynamic interactions in 2D are ultra-long-ranged, due to the logarithmic dependence on the distance r. For biomembranes, one can use the estimate~\cite{Diamant_09} $\kappa^{-1}=\eta_L h/(2\eta)$  where $\eta$   is the viscosity of the surrounding aqueous medium. Typically, $\kappa^{-1}$  is of the order of a micrometer. At separations larger than this distance, 3D effects become important for biomembranes. Our analysis applies only for relatively small membranes of micrometer size; for larger membranes the crossover to 3D hydrodynamics at long length scales needs to be taken into account.

The general expressions in eqs.~(\ref{eq:diff-gen}) and (\ref{eq:drift-gen}) hold in the 2D case as well; however, because of the logarithmic distance dependence in the Green function, hydrodynamic effects are nonlocal. Therefore, it is not possible to obtain estimates similar to those in eqs.~(\ref{eq:diff-est}) and (\ref{eq:drift-est}) for such systems. Still, some general estimates can be made.

Suppose that the concentration distribution of active inclusions has the form $c({\bf r})=c_0 \chi({\bf r})$  where $\chi({\bf r})$  is of order unity and varies on some characteristic length scale  $\ell_0 \approx \kappa^{-1}$. Substituting this into eq.~(\ref{eq:drift-gen}) and introducing the rescaled dimensionless coordinates ${\bf z}={\bf r}/\ell_0$ , we obtain
\begin{equation}
D({\bf R}_0) \approx \frac{S c_0}{\eta_m^2} \zeta_2({\bf R}_0),
\end{equation}
where the dimensionless function $\zeta_2({\bf R}_0)$  is given by
\begin{eqnarray}\label{eq:zeta2}
&&\zeta_{2}({\bf R}_0) = \frac{\Omega_{\beta \beta' \gamma \gamma'}}{32 \pi^2} \\
&&  \quad \times \int^{\prime} d{\bf z}\; \frac{\partial q_{\alpha \beta}({\bf R}_0/\ell_0-{\bf z})}{\partial z_\gamma} \frac{\partial q_{\alpha \beta'}({\bf R}_0/\ell_0-{\bf z})}{\partial z_{\gamma'}}\chi({\bf z}),\nonumber
\end{eqnarray}
and  $q_{\alpha \beta}({\bf z})=-\ln(z)\delta_{\alpha \beta}+ \hat{z}_\alpha \hat{z}_\beta$.

In a similar manner, the drift velocity in 2D systems can be estimated:
\begin{equation}
\overline{V}_\alpha({\bf R}_0) \approx \frac{S c_0}{\ell_0 \eta_m^2} \zeta_{2,\alpha}({\bf R}_0),
\end{equation}
where
\begin{eqnarray}\label{eq:xi2}
&&\xi_{2,\alpha}({\bf R}_0) = \frac{\Omega_{\beta \beta' \gamma \gamma'}}{16 \pi^2} \\
&& \quad \times \int^{\prime} d{\bf z}\; \frac{\partial^2 q_{\alpha \beta}({\bf R}_0/\ell_0-{\bf z})}{\partial z_\gamma \partial z_\delta} \frac{\partial q_{\delta \beta'}({\bf R}_0/\ell_0-{\bf z})}{\partial z_{\gamma'}} \chi({\bf z}).\nonumber
\end{eqnarray}

An important difference with the 3D case is that now the diffusion constant and the drift velocity are not determined only by the local concentration of the protein machine and its gradient. Instead, they depend on the concentration distribution over the entire 2D system through the dimensionless functions $\zeta_{2}$  and  $\xi_{2,\alpha}$. The integrals eqs.~(\ref{eq:zeta2}) and (\ref{eq:xi2}) diverge logarithmically at large distances and thus the finite size of the membrane is important.

\section{Numerical estimates}
The magnitude of a force dipole $m$ of a protein machine can be roughly estimated as $m \sim Fd_P$ , where $F$ is the force generated by the machine and $d_P$ is the linear size of the protein. Molecular motors, such as myosin or kinesin, typically generate forces about 1 pN and this can be chosen as the characteristic value for $F$. Taking the size of a protein to be about 10 nm, the force dipole can be estimated to be $m = 10^{-20}$ N m. The correlation time for force-dipole fluctuations can be taken to be the duration $t_c$ of the turnover cycle in a chemical machine. We choose a time of about $t_c \sim 10$ ms. The parameter $S$ can then be evaluated to give $S \sim m^2 t_c = 10^{-42}$ ${\rm N}^2 {\rm m}^2$ s. Note that this estimate corresponds to substrate (ATP) saturation conditions: the machine binds a new substrate molecule and enters into a new cycle immediately once the previous cycle finishes. If this condition is not satisfied, the protein machine must wait for a new ATP molecule to arrive. During this waiting period, the machine does not act as a force dipole and this will decrease the value of S.

Concentrations of active proteins inside a biological cell can vary over a large range. The highest concentrations of the order of $10^{-4}$ M are  characteristic for the enzymes involved in glycolysis. As a rough estimate, a value of $10^{-6}$ M can be chosen, corresponding to $c_0 = 10^{21} {\rm m}^{-3}$, so that the mean distance between the proteins is about 100 micrometers. Given the protein size, we choose a cutoff length of $\ell_c=20$ nm. The viscosity of water is about $10^{-3}$ Pa s. The dimensionless numerical factors in eqs.~(\ref{eq:zeta}) and (\ref{eq:xi}) can be computed and take the approximate values $\zeta=10^{-2}$ and $\xi=10^{-3}$.

With these values, the contribution (\ref{eq:diff-est}) to the diffusion coefficient due to hydrodynamic effects arising from protein machines in bulk 3D solutions is about $D = 10^{-7} {\rm cm}^2/{\rm s}$. This result should be compared with typical diffusion constants in water that can vary from about $10^{-5} {\rm cm}^2/{\rm s}$ for small molecules to $10^{-7} {\rm cm}^2/{\rm s}$ for small proteins of the size considered here in water. To estimate the magnitude of the drift velocity from eq.~(\ref{eq:drift-est}), we can take $\nabla c=\Delta c/L$  where $\Delta c$ is the concentration difference across the cell and $L=10\; \mu{\rm m}$  is the typical eukaryotic cell size. If we take $\Delta c \approx 0.1 c_0$, where again $c_0 = 10^{21} {\rm m}^{-3}$, we obtain a drift velocity of about $\overline{V}_\alpha=1 \; \mu{\rm m}/{\rm s}$.

Proceeding to lipid bilayers, we observe that their 3D viscosity $\eta_L \sim 1$ Pa s is typically a factor of $10^3$ higher than that of water. The 2D viscosity of such bilayers is $\eta_m=\eta_L h$ where $h \sim 1$ nm has been taken as the thickness of the bilayer. The magnitude $S \sim 10^{-42}\; {\rm N}^2 {\rm m}^2 {\rm s}$ may again be used for the force dipoles. Taking the mean distance between proteins to be approximately 100 nm, the 2D concentration is about $c_0 = 10^{14}\; {\rm m}^{-2}$. The dimensionless factors $\zeta_2$   and $\xi_{2, \alpha}$  depend on the concentration distribution $c({\bf r})$ of active protein inclusions over the entire membrane, according to eqs.~(\ref{eq:zeta2}) and (\ref{eq:xi2}). As rough estimates of these factors we choose $\zeta_2 = \xi_{2,\alpha}=10^{-2}$.

Given these numerical values, the hydrodynamic effects of active protein inclusions are predicted to increase the diffusion of passive particles within the membrane by about $D = 10^{-8} {\rm cm}^2/{\rm s}$ . For comparison, thermal diffusion constants for proteins in lipid bilayers are of the order of $10^{-10} {\rm cm}^2/{\rm s}$ and diffusion constants for lipids are about $10^{-8} {\rm cm}^2/{\rm s}$. The characteristic drift velocity of passive particles in lipid bilayers is estimated to be about $\overline{V}_\alpha=1 \; \mu{\rm m}/{\rm s}$ , assuming that the characteristic length for concentration variation in a membrane is about one micrometer.

These numerical estimates should be used only as rough guide to the possible magnitudes of the effects because many of the parameters may vary significantly from one system to another, or are only known poorly. For example, forces have only been measured for some molecular motors and, for protein machines that are not motors or for enzymes, they may be smaller than one pN. However, the concentrations of some proteins that behave as force dipoles may be significantly higher than the value we have assumed. For example, the enzyme phosphoglycerate kinase involved in glycolysis is present in the living cell in the concentrations up to $10^{-4}$ M, two orders of magnitude larger than our assumed value. Proteins typically contribute about $40\%$ of the mass in biomembranes and their concentrations may well be significantly higher than our 2D values. There is also some ambiguity in the choice of the cutoff length in 3D systems but its lower limit is controlled by the distance corresponding to the volume exclusion of passive and active particles. Consequently, the uncertainty in the estimates of hydrodynamic effects is high and deviations of up to two orders of magnitude from the numerical estimates given above may well be possible. Moreover, these estimates hold only under conditions where the substrate (ATP) is in excess. If the rates of machine cycling are limited by substrate supply, the effects are weaker and obviously vanish in absence of the substrate.

\section{Discussion and conclusions}
In a biological cell there are large populations of active proteins, both molecular motors and enzymes that change their conformations within catalytic cycles. In this paper we showed that when active proteins are present, either in solution or in lipid bilayers, they can substantially modify the diffusion constants of passive particles in the system. These modifications affect all passive particles, and all active proteins, even of different kinds, contribute to the effect provided they are supplied with substrate and remain active. The magnitude of the effect can be comparable to the value of thermal diffusion constants under physiological conditions. Furthermore, if protein machines are nonuniformly distributed in a cell or in a biomembrane, directed drift of passive particles, analogous to chemotaxis, can occur; however, the mechanism is completely different: all active proteins contribute towards it and all passive particles experience the drift. Drift velocities of the order of micrometers per second can be realized. The enhancement of diffusion and chemotaxis-like drift should  take place for the protein machines (enzymes) themselves as well and
 we note that enhanced diffusion of catalytically-active enzymes has been observed and is a topic of considerable current interest.~\cite{sengupta:13,sen-enz:10}

In three dimensions, hydrodynamic interactions are already long-ranged, with power-law dependence. In two dimensions they become ultra-long ranged with a logarithmic dependence on distance. Thus, the effects predicted to exist in 3D and 2D systems differ substantially. In solution, the change in the diffusion constant is determined by the local protein machine concentration and the drift velocity is controlled by its spatial gradient. In contrast,  in 2D systems such as lipid bilayers, the effects are essentially nonlocal: the change in the diffusion constant and the drift are determined by the distribution of active inclusions over the entire membrane.

Our description of how active molecular machines, through hydrodynamic interactions, influence the dynamics of passive particles was based on the equations of continuum hydrodynamics. One might question the use of such a continuum description for molecular systems. It is well established~\cite{alder:67} that hydrodynamic effects are observable on very small scales of tens of solvent particle diameters. They persist in spite of strong fluctuations and their presence is signalled in the long-time tails of velocity correlation functions or even in the transport properties of polymers. Our use of continuum equations is restricted to rather long scales so that the main conclusions of our study should be robust.

Our analysis has focused on general aspects of the phenomena and is not intended to address the full complexity seen in biological systems. The cell is crowded with obstacles of various kinds and these crowding elements could modify the hydrodynamic behavior. Nevertheless, the results suggest a physical picture of the dynamics in the cell that differs from one based solely on thermal motions. Whenever active proteins are present, they not only perform their varied biological functions, but also generate hydrodynamic fields that span the entire cell. Such non-thermal fluctuating fields affect the diffusion of particles and lead to their directed drift.  There is another important effect, which we have not analyzed, but is worth noting: because such fluctuating fields represent non-thermal noise, they can be rectified and work or energy can be extracted from them. Through hydrodynamic interactions, active inclusions supply power in a distributed way within the cell or a biological membrane.

\section{Acknowledgements}
\label{sec:acknowledge}
This work of RK was supported by a grant from the Natural Sciences and Engineering Research Council of Canada. Financial support from the Volkswagen Foundation for AM and from the Deutsche Forschungsgemeinschaft for RK is gratefully acknowledged.
%

\end{document}